\documentclass[twocolumn,amsmath,amssymb,superscriptaddress]{revtex4-1}

\usepackage{lineno}
\usepackage{xspace}
\usepackage{calc}
\usepackage{tikz}
\usepackage{siunitx}

\newcommand{\ie}{i.e.\@\xspace}

\newcommand{\ud}{\mathrm{d}}
\newcommand{\bra}{\left\langle}
\newcommand{\ket}{\right\rangle}

\newcommand{\im}{\operatorname{Im}}
\newcommand{\opv}{\mathcal{V}}
\newcommand{\opg}{\mathcal{G}}
\newcommand{\opsigma}{\mathcal{S}}

\newlength{\diaglength}
\newcounter{diagcount}
\newcommand{\diag}[2]{
   \setlength{\diaglength}{#1mm}
   \raisebox{1.2\diaglength}{\begin{tikzpicture}[scale=0.12]#2\end{tikzpicture}}
}
\newcommand{\particule}[1]{\draw [fill=white] (#1,0) circle (1) ;}
\newcommand{\gr}[2]{\draw (#1,0) -- (#2,0) ;}

\newcommand{\correldeux}[2]{
   \setcounter{diagcount}{#2}
   \addtocounter{diagcount}{-#1}
   \setcounter{diagcount}{\thediagcount*\real{0.5}}
   \draw [dashed] (#1,0) arc (180:0:\thediagcount) ;
}
\newcommand{\correltrois}[3]{
   \setcounter{diagcount}{#2}
   \draw [dashed] (\thediagcount,6) -- (#1,0) ;
   \draw [dashed] (\thediagcount,6) -- (#2,0) ;
   \draw [dashed] (\thediagcount,6) -- (#3,0) ;
}

\begin{document}

\title{Multiple scattering theory in one dimensional space and time dependent disorder: Average field}

\author{Alexandre Selvestrel}
\affiliation{Institut Langevin, ESPCI Paris, Université PSL, CNRS, F-75005 Paris, France}
\author{Julia Rocha}
\affiliation{Institut Langevin, ESPCI Paris, Université PSL, CNRS, F-75005 Paris, France}
\author{Rémi Carminati}
\affiliation{Institut Langevin, ESPCI Paris, Université PSL, CNRS, F-75005 Paris, France}
\affiliation{Institut d’Optique Graduate School, Université Paris-Saclay, F-91127 Palaiseau, France}
\author{Romain Pierrat}
\affiliation{Institut Langevin, ESPCI Paris, Université PSL, CNRS, F-75005 Paris, France}

\begin{abstract}
   We theoretically study the propagation of light in one-dimensional space- and time-dependent disorder.
   The disorder is described by a fluctuating permittivity $\epsilon(x,t)$ exhibiting short-range correlations
   in space and time, without cross correlation between them. Depending on the illumination conditions, we
   show that the intensity of the average field decays exponentially in space or in time, with characteristic length or time defining
   the scattering mean-free path $\ell_s$ and the scattering mean-free time $\tau_s$. In the weak scattering regime, we provide
   explicit expressions for $\ell_s$ and $\tau_s$, that are checked against rigorous numerical simulations.
\end{abstract}

\maketitle

\section{Introduction}
% ====================

Light (or more generally wave) propagation in spatially disordered media has been an active topic for many decades,
stimulated by basic questions in fundamental physics and by a large number of applications.  On the fundamental side,
the existence of Anderson localization for different kinds of waves is an emblematic example, among many other questions
in mesoscopic physics~\cite{SHENG-2006}. On the applied side, imaging and sensing~\cite{SEBBAH-2001} or light control in
complex materials~\cite{GIGAN-2022} are highly developed research themes. The basic concepts and theoretical tools for
modeling light propagation in spatially disordered media are known to a large extent~\cite{CARMINATI-2021-1}.

Beyond spatial modulation of the medium, there has recently been a surge in research on propagation of different kinds
of waves in time-dependent media, including electromagnetic~\cite{CALOZ-2020},
optical~\cite{LUSTIG-2018,SHARABI-2021,SAHA-2023,TIROLE-2023}, acoustic~\cite{ZANGENEH-NEJAD-2019} or water
waves~\cite{BACOT-2016,BACOT-2019}.  This emerging field opens new perspectives in terms of applications. For example,
periodic space-time metamaterials offer new degrees of freedom for wave
control~\cite{AKBARZADEH-2018,PACHECO-PENA-2020,SHARABI-2022}.  It also stimulates the development of appropriate
theories, in an area that has been largely unexplored so far.  For example, some of us have highlighted the atypical
behavior of wave propagation in a time-varying disorder, showing that the average energy of the field grows exponentiall
at long times~\cite{CARMINATI-2021}, providing a theoretical support to observations based on numerical
simulations~\cite{SHARABI-2021} or experiments~\cite{APFFEL-2022}. Another recent study has focused on the role of
correlations in time disorder in providing innovative optical properties~\cite{KIM-2023}. These bricks contribute to
the development of theories of wave propagation in time-varying disordered media, which remains a widely open topic.

In this article, we address the question of light propagation in a medium exhibiting both space and time disorders. To
start with a simple model, we consider a one-dimensional space disorder combined to a time modulation, resulting in a
medium described by a fluctuating dielectric function $\epsilon(x,t)$ considered to be a random variable, with $x$ and
$t$ the space and time coordinates, respectively. We assume that the medium exhibits short-range correlations in both
space and time, without cross correlation between them.  The main objective is to develop a theory for the average
field (or intensity) proving the existence of a scattering mean-free path $\ell_s$ and a scattering mean-free time
$\tau_s$, and to provide explicit expressions in the weak scattering regime.  The paper is organized as follows: In
Sec.~\ref{theory}, we develop the theory that extends the standard multiple scattering theory to a situation with both
space and time disorders. We provide expressions for $\ell_s$ and $\tau_s$ using a perturbative approach. In
Sec.~\ref{gauss}, we consider the particular case of disorder with a gaussian correlation in space
and time, and show that the expressions of the mean-free path and mean-free time are in full agreement with numerical
simulations performed without approximations.

\section{Multiple scattering theory for space-time disorder}\label{theory}
% ==========================================================

In this section we build a theory to compute the average electric field, from which we will define $\ell_s$ and $\tau_s$, and derive
their explicit expressions. To proceed, we generalize the standard multiple scattering theory to account for
space-time disorder. The interested reader can find detailed presentations of multiple scattering theory in
various textbooks~\cite{RYTOV-1989,SHENG-2006,MONTAMBAUX-2007,CARMINATI-2021-1}. In a medium with one-dimensional
space-time disorder described by a random dielectric function $\epsilon(x,t)$, an electric field linearly polarized along the $y$-direction obeys the
equation
\begin{equation}\label{eq:wave_equation}
   -\frac{\partial^2 E(x,t)}{\partial x^2}+\frac{1}{c^2}\frac{\partial^2}{\partial t^2}\left[\epsilon(x,t)E(x,t)\right]=S(x,t) \, ,
\end{equation}
which is easily derived form Maxwell's equations. Here $E(x,t)$ is the real amplitude of the field in the time domain,
$c$ is the speed of light in vacuum, and $S(x,t)$ is a source term that we do not need to specify. 
It is interesting to note that in Eq.~\eqref{eq:wave_equation} the dielectric function remains within the time
derivative operator, which has important consequences as will be seen later. This is a feature of scattering problems involving 
two types of disorder, for example with both permittivity and permeability disorders~\cite{BORN-1999}, or in acoustics with
mass density and compressibility disorders~\cite{BAYDOUN-2015}. We also note that working with the displacement field
$D$, as in Ref.~\cite{CARMINATI-2021}, does not simplify the equation when space and time disorders coexist.

\subsection{Lippmann-Schwinger equation}
% --------------------------------------

The first step of the derivation consists in defining a homogeneous reference (or background) medium with permittivity $\epsilon_b$. The
reference field $E_b$ in this medium satisfies
\begin{equation}\label{eq:wave_equation_reference}
   -\frac{\partial^2 E_b(x,t)}{\partial x^2}+\frac{\epsilon_b}{c^2}\frac{\partial^2 E_b(x,t)}{\partial t^2}=S(x,t).
\end{equation}
The choice of $\epsilon_b$ will be specified later, with the constraint that it should be close to the typical value of
$\epsilon(x,t)$ to ensure the accuracy of the perturbative approach. 

Subtracting Eq.~\eqref{eq:wave_equation_reference} from Eq.~\eqref{eq:wave_equation} leads to
\begin{multline}\label{eq:wave_equation_scattered}
   -\frac{\partial^2 E_s(x,t)}{\partial x^2}+\frac{\epsilon_b}{c^2}\frac{\partial^2 E_s(x,t)}{\partial t^2}
      =\frac{\epsilon_b}{c^2}\frac{\partial^2 E(x,t)}{\partial t^2}
\\
      -\frac{1}{c^2}\frac{\partial^2}{\partial t^2}\left[\epsilon(x,t)E(x,t)\right]
\end{multline}
where $E_s=E-E_b$ is the scattered field. Equation~\eqref{eq:wave_equation_scattered} shows that the scattered field can be 
seen as a field propagating in the reference medium and caused by a complex source term given by the right-hand side. 
We now introduce the Green function $G_b$ defined as the solution to
\begin{multline}\label{eq:wave_equation_reference_green}
   -\frac{\partial^2 G_b(x-x',t-t')}{\partial x^2}+\frac{\epsilon_b}{c^2}\frac{\partial^2 G_b(x-x',t-t')}{\partial t^2}
\\
      =\delta(x-x')\delta(t-t')
\end{multline}
where $\delta$ is the Dirac delta function, satisfying Sommerfeld's radiation condition in space (or equivalently causality in time).
The Green function can be understood as the electric field radiated in the reference medium by a point source emitting
an infinitely short pulse. Its detailed calculation is given in App.~\ref{green_calculation}. In the Fourier domain,
the Green function is
\begin{multline}\label{eq:green_fourier}
   G_b(k,\omega)=\operatorname{PV}\left[\frac{1}{k^2-\omega^2/v^2}\right]
      +\frac{i\pi}{2k}\delta\left(k-\frac{\omega}{v}\right)
\\
      -\frac{i\pi}{2k}\delta\left(k+\frac{\omega}{v}\right)
\end{multline}
where $v=c/\sqrt{\epsilon_b}$ is the phase velocity in the reference medium and $\operatorname{PV}$ stands for the
Cauchy principal value operator. In the following, we will consider two different problems: (1)~the
evolution of the wave in space for a monochromatic incident beam, and (2) the evolution of the wave in time for an
incident pulse with a fixed wave number. Expressions for the reference Green function in the $(x,\omega)$ and in the
$(k,t)$ domains are thus required for problems (1) and (2), respectively. They are given by
\begin{equation}\label{eq:green_time_space}
   \left\{\begin{aligned}
      G_b(x,\omega) & =\frac{i}{2k_b}\exp\left(ik_b|x|\right),
   \\
      G_b(k,t) & =\frac{\operatorname{H}(t)v^2}{\omega_b}\sin\left(\omega_b t\right),
   \end{aligned}\right.
\end{equation}
where $\operatorname{H}$ is the Heaviside step function, $k_b=\omega/v$ and $\omega_b=kv$. It is important to note that
the observed asymmetry between space and time arises from the different boundary conditions in both cases. In
space, we have considered the Sommerfeld's radiation condition which states that the field should propagate towards
$+\infty$ (respectively $-\infty$) for $x>0$ (respectively $x<0$). In time, this condition is equivalent to causality
which translates into the presence of the $\operatorname{H}$ operator.

Equation~\eqref{eq:wave_equation_scattered} together with the definition of the Green function $G_b$ allows us to write
the scattered field $E_s$ in the integral form
\begin{multline}
   E_s(x,t)=-\iint G_b(x-x',t-t')\frac{\partial^2\left[\epsilon(x',t')-\epsilon_b\right]E(x',t')}{c^2\partial t'^2}
\\\times
   \ud x'\ud t'.
\end{multline}
The total field $E=E_b + E_s$ obeys the integral equation
\begin{multline}\label{eq:lippmann-schwinger}
   E(x,t)=E_b(x,t)-\iint G_b(x-x',t-t')
\\\times
   \frac{\partial^2\left[\epsilon(x',t')-\epsilon_b\right]E(x',t')}{c^2\partial t'^2}\ud x'\ud t' \, ,
\end{multline}
known as the Lippmann-Schwinger equation. This equation is the elementary building block of multiple scattering theory. 
For further developments, it will prove useful to manipulate formal operator expressions. To that end, we define 
\begin{equation}
   \left\{\begin{aligned}
      \opv & =-\frac{\partial^2\left[\epsilon(x',t')-\epsilon_b\right]\bullet}{c^2\partial t^2},
      \\
      \opg_b & =\iint G_b(x-x',t-t')\bullet\ud x'\ud t',
   \end{aligned}\right.
\end{equation}
where the bullets are to be uderstood as the quantity on which the operator acts. In this formalism, the
Lippmann-Schwinger equation~\eqref{eq:lippmann-schwinger} can be rewritten as
\begin{equation}\label{eq:lippmann-schwinger_formal}
   E=E_b+\opg_b\opv E.
\end{equation}
We emphasize that the main difference with the usual Lippmann-Schwinger equation appearing in standard multiple scattering
theory is the operator character of the scattering potential $\opv$.

\subsection{Born series and Dyson equation}
% -----------------------------------------

In order to estimate the field averaged over an ensemble of realizations of disorder (\ie, of the random variable
$\epsilon(x,t,)$), we first expand Eq.~\eqref{eq:lippmann-schwinger_formal} in the form
\begin{equation}\label{eq:born_series}
   E=E_b+\opg_b\opv E_b+\opg_b\opv\opg_b\opv E_b+\opg_b\opv\opg_b\opv\opg_b\opv E_b+\ldots
\end{equation}
which is known as the Born series. Performing a statistical ensemble average, we find that
\begin{equation}\label{eq:born_series_average}
   \bra E\ket=E_b+\opg_b\bra \opv\ket E_b+\opg_b\bra \opv\opg_b\opv\ket E_b+\opg_b\bra \opv\opg_b\opv\opg_b\opv\ket E_b+\ldots
\end{equation}
where $\bra\ldots\ket$ denotes the average value. 
The problem now reduces to the computation of terms of the form $\bra\opv(\opg_b\opv)^n\ket$. Let us first consider
the second order term (\ie, $n=1$). We define the connected part of the correlation function of the potentiel by
\begin{equation}\label{eq:connection}
   \bra \opv\opg_b\opv\ket=\bra \opv\ket\opg_b\bra\opv\ket+\bra \opv\opg_b\opv\ket_c.
\end{equation}
This corresponds to a splitting of the correlation function into a factorizable and a non-factorizable (connected) part.
Similar splittings for more complicated terms would require relatively heavy writing. A convenient way to manipulate such
expressions is to use diagrams. For Eq.~\eqref{eq:connection}, we write
\begin{equation}\label{eq:connection_diag}
   \bra \opv\opg_b\opv\ket=
   \diag{0}{
      \gr{1}{7}
      \particule{1}
      \particule{7}
   }
   +
   \diag{0}{
      \gr{1}{7}
      \correldeux{1}{7}
      \particule{1}
      \particule{7}
   }
\end{equation}
where circles, solid lines and dashed lines represent scattering events (interactions with the scattering potential),
Green functions of the reference medium, and connections (non-factorizable part of the correlation function),
respectively. Using diagrams, the third-order case ($n=2$) becomes
\begin{multline}\label{eq:connection_bis_diag}
   \bra \opv\opg_b\opv\opg_b\opv\ket=
   \diag{0}{
      \gr{1}{7}
      \gr{7}{13}
      \particule{1}
      \particule{7}
      \particule{13}
   }
   +
   \diag{0}{
      \gr{1}{7}
      \gr{7}{13}
      \correldeux{1}{7}
      \particule{1}
      \particule{7}
      \particule{13}
   }
   +
   \diag{0}{
      \gr{1}{7}
      \gr{7}{13}
      \correldeux{7}{13}
      \particule{1}
      \particule{7}
      \particule{13}
   }
\\
   +
   \diag{0}{
      \gr{1}{7}
      \gr{7}{13}
      \correldeux{1}{13}
      \particule{1}
      \particule{7}
      \particule{13}
   }
   +
   \diag{0}{
      \gr{1}{7}
      \gr{7}{13}
      \correltrois{1}{7}{13}
      \particule{1}
      \particule{7}
      \particule{13}
   }
\end{multline}
and similarly for higher-order terms.

The key idea to obtain an equation for the average field consists in defining a new operator $\opsigma$
containing all non-factorizable terms, \ie
\begin{equation}\label{eq:sigma}
   \opsigma=
   \diag{0}{
      \particule{1}
   }
   +
   \diag{0}{
      \gr{1}{7}
      \correldeux{1}{7}
      \particule{1}
      \particule{7}
   }
   +
   \diag{0}{
      \gr{1}{7}
      \gr{7}{13}
      \correldeux{1}{13}
      \particule{1}
      \particule{7}
      \particule{13}
   }
   +
   \diag{0}{
      \gr{1}{7}
      \gr{7}{13}
      \correltrois{1}{7}{13}
      \particule{1}
      \particule{7}
      \particule{13}
   }
   +\ldots
\end{equation}
With this definition, Eq.~\eqref{eq:born_series_average} can be factorized in the form
\begin{equation}\label{eq:dyson}
   \bra E\ket=E_b+\opg_b\opsigma\bra E\ket
\end{equation}
which is known as the Dyson equation. Equation~\eqref{eq:dyson} is exact and all the complexity of
the multiple scattering problem lies in the closed form of the equation and in the operator $\opsigma$. 
In order to define the scattering mean-free path and time, and to derive explicit expressions, we need to simplify this operator.
To that end, let us consider the first term corresponding to a single scattering event. Applying the operator to the
average field leads to
\begin{equation}\label{eq:sigma_first_order}
   \opsigma^{(1)}\bra E\ket
      =-\frac{1}{c^2}\frac{\partial^2}{\partial t^2}\left[\bra\epsilon(x,t)-\epsilon_b\ket\bra E(x,t)\ket\right].
\end{equation}
We now need to make a choice for the reference medium. Taking $\epsilon_b=\bra\epsilon(x,t)\ket$
ensures the accuracy of the pertubation method that we will use, by implying a vanishing first order in the 
perturbative expansion. Indeed, by defining the fluctuating part of the permittivity by
$\delta\epsilon(x,t)=\epsilon(x,t)-\bra\epsilon(x,t)\ket=\epsilon(x,t)-\epsilon_b$ we find that
\begin{equation}
   \opsigma^{(1)}\bra E\ket
      =-\frac{1}{c^2}\frac{\partial^2}{\partial t^2}\left[\bra\delta\epsilon(x,t)\ket\bra E(x,t)\ket\right]=0.
\end{equation}
For the second order term, we obtain
\begin{multline}\label{eq:sigma_second_order}
   \opsigma^{(2)}\bra E\ket
      =\frac{1}{c^4}\iint \frac{\partial^2}{\partial t^2}\left\{\bra\delta\epsilon(x,t)G_b(x-x',t-t')
\vphantom{\frac{\partial^2}{\partial t'^2}}\right.\right.\\\times\left.\left.
         \frac{\partial^2}{\partial t'^2}\left[\delta\epsilon(x',t')\bra E(x',t')\ket\right]\ket\right\}\ud x'\ud t'
\end{multline}
where we have used the relationship
$\bra\delta\epsilon(x,t)\delta\epsilon(x',t')\ket_c=\bra\delta\epsilon(x,t)\delta\epsilon(x',t')\ket$, generally
referred to as the correlation function of disorder. The first second-order time derivative relates to the variable $t$ and
thus applies only to $\delta\epsilon(x,t)G_b(x-x',t-t')$. The second derivation relates to $t'$ and applies to
$\delta\epsilon(x',t')\bullet$. It will prove useful to perform a double integration by parts. Assuming that the 
correlation function of disorder vanishes at long times, Eq.~\eqref{eq:sigma_second_order} reduces to
\begin{multline}\label{eq:sigma_second_order_simplified}
   \opsigma^{(2)}\bra E\ket
      =\frac{1}{c^4}\iint \bra\frac{\partial^2}{\partial t^2}\left[\delta\epsilon(x,t)
         \frac{\partial^2}{\partial t'^2}G_b(x-x',t-t')\right]
\right.\\\times\left.\vphantom{\frac{\partial^2}{\partial t^2}}
         \delta\epsilon(x',t')\ket\bra E(x',t')\ket\ud x'\ud t'.
\end{multline}
Similar transformations can be performed on the higher order terms in $\opsigma$, but are not written here since they will not be
useful in practice. Finallly, $\opsigma$ can be written as
\begin{equation}
   \opsigma=\iint \Sigma(x,x',t,t')\bullet\ud x'\ud t'
\end{equation}
where $\Sigma$ is the self-energy and is here a simple multiplicative function (not an operator). Using the self-energy, the
average field can be written
\begin{multline}\label{eq:dyson_simplified}
   \bra E(x,t)\ket=E_b(x,t)+\idotsint G_b(x-x',t-t')
\\\times
   \Sigma(x',x'',t',t'')\bra E(x'',t'')\ket\ud x'\ud x''\ud t'\ud t'' \, ,
\end{multline}
which is the integral form of the Dyson equation.

\subsection{Weak-scattering regime}
% ---------------------------------

To derive expressions for the scattering mean-free path and time, we now consider the particular case of a source term
of the form $S(x,t)=\delta(x)\delta(t)$ in an infinite medium. In this case, $E_b(x,t)=G_b(x,t)$ and $\bra E(x,t)\ket=\bra
G(x,t)\ket$, with $G$ the Green function of the medium in the presence of disorder. We assume statistical homogeneity
in space and time, such that $\Sigma(x',x'',t',t'')$ only depends on $x'-x''$ and $t'-t''$. In these conditions, 
Eq.~\eqref{eq:dyson_simplified} reduces to
\begin{multline}\label{eq:dyson_green}
   \bra G(x,t)\ket=G_b(x,t)+\idotsint G_b(x-x',t-t')
\\\times
   \Sigma(x'-x'',t'-t'')\bra G(x'',t'')\ket\ud x'\ud x''\ud t'\ud t''.
\end{multline}
This equation can be solved by performing a space-time Fourier transform, which leads to
\begin{equation}\label{eq:dyson_fourier}
   \bra G(k,\omega)\ket=G_b(k,\omega)+G_b(k,\omega)\Sigma(k,\omega)\bra G(k,\omega)\ket \, ,
\end{equation}
or equivalently
\begin{equation}
   \frac{1}{\bra G(k,\omega)\ket}=\frac{1}{G_b(k,\omega)}-\Sigma(k,\omega)
      =k^2-\frac{\omega^2}{v^2}-\Sigma(k,\omega).
\end{equation}
This expression of the average Green function will be used to derive expressions for the
scattering mean-free path $\ell_s$ and scattering mean-free time $\tau_s$. 
We start by considering a monochromatic source term, oscillating at a frequency $\omega$,
and we focus on the spatial behavior of the average field given by
\begin{equation}\label{eq:green_spatial}
   \bra G(x,\omega)\ket=\int_{-\infty}^{+\infty}\frac{e^{ikx}}{k^2-k_b^2-\Sigma(k,\omega)}\frac{\ud k}{2\pi}.
\end{equation}
The computation of this inverse Fourier transform requires additional hypotheses. Considering the 
weak-scattering regime defined by the condition $|\Sigma(k,\omega)|\ll k_b^2$, the self-energy has a significant contribution only
when $k \simeq \pm k_b$. Assuming that the disorder is statistically isotropic, we also
have $\Sigma(k,\omega)=\Sigma(-k,\omega)$. As a result, the self-energy can be taken on-shell for $k=k_b$ in
Eq.~\eqref{eq:green_spatial}. Under this assumption, Eq.~\eqref{eq:green_spatial} becomes
\begin{equation}\label{eq:green_spatial_on_shell}
   \bra G(x,\omega)\ket\approx\int_{-\infty}^{+\infty}\frac{e^{ikx}}{k^2-k_b^2-\Sigma(k_b,\omega)}\frac{\ud k}{2\pi}.
\end{equation}

% Contours average
\begin{figure*}[!htb]
   \centering
   \includegraphics[width=\linewidth]{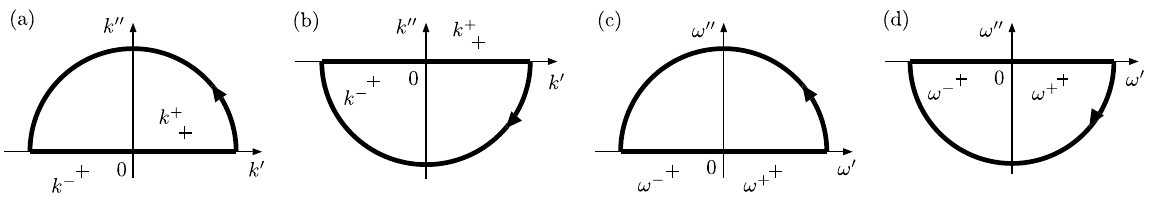}
   \caption{Integration contours used to compute the average Green function. (a) and (b) are used for the
   inverse Fourier transform for positive and negative positions, respectively. (c) and (d) are used for the inverse Fourier
   transform for negative and positive times, respectively. In these representations, we have assumed
   $\im\Sigma(k_b,\omega)>0$ for (a) and (b) and $\im\Sigma(k,\omega_b)>0$ for (c) and (d), as explained in the
   main text.}
   \label{fig:contours_average}
\end{figure*}

In order to compute the integral, we apply the residue theorem. For $x>0$, we use the contour plotted in
Fig.~\ref{fig:contours_average}\,(a). The semicircle in the upper plane is chosen in order to apply Jordan's lemma. 
The poles are $k^{\pm}=\pm\sqrt{k_b^2+\Sigma(k_b,\omega)}$. Assuming that $\im\Sigma(k_b,\omega)>0$, which will be
justified below, we obtain
\begin{equation}
   \bra G(x>0,\omega)\ket=\frac{ie^{ik^+x}}{k^+-k^-}=\frac{ie^{ik^+x}}{2k^+}.
\end{equation}
For $x<0$, we use the contour presented in Fig.~\ref{fig:contours_average}\,(b) and we obtain
\begin{equation}
   \bra G(x<0,\omega)\ket=\frac{ie^{-ik^+x}}{k^+-k^-}=\frac{ie^{-ik^+x}}{2k^+}
\end{equation}
which finally leads to 
\begin{equation}\label{eq:green_average_spatial}
   \bra G(x,\omega)\ket=\frac{ie^{ik_e|x|}}{2k_e} \, ,
\end{equation}
where $k_e=k^+$. We clearly see from Eqs.~\eqref{eq:green_time_space} and \eqref{eq:green_average_spatial} that the
average field propagates in a homogeneous effective medium with an effective wavevector $k_e$. It is convenient to split
$k_e$ into its real and imaginary parts. We write $k_e=k_r+i/(2\ell_s)$, with $\ell_s$ the scattering mean-free path,
and $k_r=n_r\omega/c$, which defines the real part $n_r$ of the effective refractive index of the medium. The intensity
of the average field is then given by
\begin{equation}\label{eq:intensity_spatial}
   \left|\bra G(x,\omega)\ket\right|^2=\frac{e^{-|x|/\ell_s(\omega)}}{4\left|k_e\right|^2}
\end{equation}
which will be the expression used later for comparison with numerical simulations.
In the weak scattering regime, we have
\begin{equation}\label{eq:scattering_mean_free_path}
   \ell_s(\omega)=\frac{k_b}{\im\Sigma(k_b,\omega)}.
\end{equation}
We also note that the approximation $k_r\simeq k_b$ holds in the weak-scattering regime. A more refined expression
would involve the real part of the self-energy.

We now turn to the illumination by a pulse source term with a fixed $k$-vector, and we focus on the temporal evolution of the average
Green function, which is given by
\begin{equation}\label{eq:green_temporal}
   \bra G(k,t)\ket=\int_{-\infty}^{+\infty}\frac{e^{-i\omega t}}{\omega_b^2/v^2-\omega^2/v^2-\Sigma(k,\omega)}\frac{\ud\omega}{2\pi}.
\end{equation}
The weak-scattering regime amounts to assuming that $|\Sigma(k,\omega)|\ll\omega_b^2/v^2$. Under this
assumption, the self-energy takes significant values for $\omega \simeq \pm \omega_b$. For statistically isotropic disorder, such that
$\Sigma(k,\omega)=\Sigma(-k,\omega)$, and making use of the fact that $\Sigma(x,t)$ is real valued, we find that
\begin{multline}
   \Sigma(k,-\omega)=\int_{-\infty}^{+\infty}\Sigma(x,t)e^{-ikx-i\omega t}\ud x\ud t
\\
      =\left[\int_{-\infty}^{+\infty}\Sigma(x,t)e^{ikx+i\omega t}\ud x\ud t\right]^*
\\
      =\left[\int_{-\infty}^{+\infty}\Sigma(x,t)e^{-ikx+i\omega t}\ud x\ud t\right]^*
      =\Sigma(k,\omega)^*.
\end{multline}
As a result, the self-energy can be replaced by $\Sigma(k,\omega_b)^*$ in Eq.~\eqref{eq:green_temporal} in the vicinity of
$-\omega_b$, and by $\Sigma(k,\omega_b)$ in the vicinity of $\omega_b$. This is the counterpart of the on-shell
approximation in the frequency domain. In order to compute the integral, we now make use of the residue theorem.
For $t<0$, we use the contour in Fig.~\ref{fig:contours_average}\,(c). The poles are 
$\omega^-=-v\sqrt{k^2-\Sigma^*(k,\omega_b)}$ and $\omega^+=v\sqrt{k^2-\Sigma(k,\omega_b)}$. If $\im\Sigma(k,\omega_b)>0$,
we get $\bra G(k,t<0)\ket=0$. This is the signature of causality in the time-domain Green function. For $t>0$, considering the contour in
Fig.~\ref{fig:contours_average}\,(d), we find that
\begin{equation}
   \bra G(k,t>0)\ket=iv^2\left[
         \frac{e^{-i\omega^+t}}{\omega^+-\omega^-}
        +\frac{e^{-i\omega^-t}}{\omega^--\omega^+}
      \right] .
\end{equation}
Defining $\omega_e=\omega_r-i/(2\tau_s)$, with $\tau_s$ the scattering mean-free time, we finally obtain
\begin{equation}\label{eq:green_average_temporal}
   \bra G(k,t)\ket=\frac{\operatorname{H}(t)v^2}{\omega_r}\sin(\omega_r t)e^{-t/[2\tau_s(k)]}
\end{equation}
for the average field. To wash out the rapid oscillations in the intensity, we square this expression and perform 
a time average over a window with width $T$ such that $2\pi/\omega_r \ll T\ll \tau_s$. This leads to
\begin{equation}\label{eq:intensity_temporal}
   \overline{\bra G(k,t)\ket^2}=\frac{\operatorname{H}(t)v^4}{2\omega_r^2}e^{-t/\tau_s(k)} \, ,
\end{equation}
which will be the expression used for comparison with numerical simulations.
Under the weak-scattering approximation, the scattering mean-free time is 
\begin{equation}\label{eq:scattering_mean_free_time}
   \tau_s(k)=\frac{k^2}{\omega_b\im\Sigma(k,\omega_b)}.
\end{equation}
We note that $\omega_r \simeq \omega_b$ in the weak-scattering regime. 
We also stress that having $\im\Sigma(k,\omega_b)<0$ is not possible since this would lead to a non vanishing
average Green function for $t<0$, thus violating causality. 

In summary, Eqs~\eqref{eq:scattering_mean_free_path} and \eqref{eq:scattering_mean_free_time} show that it is 
possible to define a scattering mean-free path $\ell_s$ and a scattering mean-free time $\tau_s$ for a space and time
dependent disorder. The reason for this is that self-energy $\Sigma$ is a simple multiplicative function, even when the
scattering potential is an operator. Moreover, we clearly see from the Dyson equation~\eqref{eq:dyson_fourier} that there 
is no change in frequency or wavevector during propagation of the average field. For a
monochromatic source at frequency $\omega$, this means that the average field propagates at $\omega$ and the scattering mean-free path
can be defined for a fixed frequency $\omega$. Similarly, for a source at a fixed wavevector $k$, the average field
evolves at the same $k$ and the scattering mean-free time can be defined for this fixed wavevector $k$. This behavior
is typical of an average (or ballistic) field, and is observed for example in dynamic multiple scattering (or diffusing-wave 
spectroscopy) where the Doppler shift vanishes for the average field~\cite{PIERRAT-2008-1}.

\section{Disorder with a Gaussian correlation in space and time}\label{gauss}
% ===================================================================================

To get explicit expressions for the scattering mean-free path $\ell_s$ and mean-free time $\tau_s$, we need to define a
specific model of disorder. A canonical choice is that of a disorder with Gaussian correlation in both space and time,
which allows to derive analytical expressions that can be easily compared to numerical simulations. This comparison is a
relevant test of validity of the pertubation theory developed above.

\subsection{Practical expressions for $\ell_s$ and $\tau_s$}
% ----------------------------------------------------------

In the weak-scattering regime, we can derive expressions for $\ell_s$ and $\tau_s$ restricted to the leading term
in the perturbative expansion of the self-energy. The self-energy reads as
\begin{multline}\label{eq:sigma_second_order_explicit}
      \Sigma(x-x',t-t')=\frac{1}{c^4} \frac{\partial^2}{\partial t^2}\left\{\bra\delta\epsilon(x,t)
         \frac{\partial^2}{\partial t'^2}\left[G_b(x-x',t-t')\right]
\right.\right.\\\times\left.\left.\vphantom{\frac{\partial^2}{\partial t^2}}
         \delta\epsilon(x',t')\ket\right\} \, ,
\end{multline}
which can be reorganized in the form
\begin{multline}\label{eq:sigma_second_order_correl}
      \Sigma(x-x',t-t')=\frac{1}{c^4} \frac{\partial^2}{\partial t^2}\left\{
         \frac{\partial^2}{\partial t'^2}\left[G_b(x-x',t-t')\right]
\right.\\\times\left.\vphantom{\frac{\partial^2}{\partial t^2}}
         \bra\delta\epsilon(x,t)\delta\epsilon(x',t')\ket
      \right\} \, , 
\end{multline}
in which the correlation function of disorder appears explicitly. We now assume that this correlation function
factorizes into two components, in the form
\begin{equation}\label{eq:correlation}
   \bra\delta\epsilon(x,t)\delta\epsilon(x',t')\ket=\alpha(x-x')\beta(t-t') \, ,
\end{equation}
meaning that short-range correlations may exist in the space or time dependence, with cross space-time
correlations excluded. Plugging this expression into Eq.~\eqref{eq:sigma_second_order_correl}, and taking the Fourier transform, leads to
\begin{equation}\label{eq:sigma_fourier_explicit}
   \Sigma(k,\omega)=\frac{\omega^2}{c^4}\int (\omega-\omega')^2\mathcal{A}(k,\omega-\omega')\beta(\omega')\frac{\ud\omega'}{2\pi}
\end{equation}
where $\mathcal{A}(k,\omega)$ is the Fourier transform of $\mathcal{A}(x,t)=G_b(x,t)\alpha(x)$. We note that for a pure
static disorder, with $\beta(t-t')=1$, we would recover the standard result involving the spatial correlation function
of disorder and the Green function, namely $\beta(\omega')=2\pi\delta(\omega')$ and $\Sigma(k,\omega)=\omega^4/c^4\mathcal{A}(k,\omega)$~\cite{VYNCK-2023}.

The gaussian-correlated disorder model amounts to considering that
\begin{equation}\label{eq:disorder_correlation_functions}
   \left\{\begin{aligned}
      \alpha(x-x') & =A\exp\left[-\frac{(x-x')^2}{2\ell^2}\right],
   \\
      \beta(t-t') & =B\exp\left[-\frac{(t-t')^2}{2\tau^2}\right]
   \end{aligned}\right.
\end{equation}
where $\ell$ and $\tau$ are the correlation length and time of disorder, respectively, and $A$ and $B$ are amplitudes of the
correlation functions. With this model, we find that
\begin{widetext}
   \begin{multline}\label{eq:sigma_gauss}
      \im\Sigma(k,\omega)= \frac{AB\ell\tau\omega^4 v^3}{8c^4\left(\ell^2+\tau^2v^2\right)^{3/2}}e^{-\eta^2}
         \left\{
            \sqrt{2\pi}\left(\omega\tau^2v-k\ell^2\right)e^{\xi^2}
            +2\sqrt{2\pi}\left(\omega\tau^2v+k\ell^2\right)e^{2k\omega\ell^2/v+\xi^2}
   \right.\\\left.
            +2\sqrt{\ell^2+\tau^2v^2}
            +\sqrt{2\pi}\left(k\ell^2-\omega\tau^2v\right)\operatorname{erf}(\xi)e^{\xi^2}
         \right\}
         -\frac{AB\ell\omega^4\tau v^3}{4c^4\ell^2\sqrt{\pi}\left(\omega+kv\right)}e^{-\eta^2}
         \left\{
            \omega G_{0,0}^{2,1}\left(-\xi,\frac{1}{2}\left|\begin{matrix}1\\1/2,1\end{matrix}\right.\right)
   \right.\\\left.
            +\frac{2v^2}{\ell^2\left(\omega+kv\right)}
               G_{0,0}^{2,1}\left(-\xi,\frac{1}{2}\left|\begin{matrix}1\\1,3/2\end{matrix}\right.\right)
         \right\}
   \end{multline}
   where
   \begin{equation}
      \eta=\frac{\ell(\omega+kv)}{\sqrt{2}v}
      \quad;\quad
      \xi=\frac{\ell^2(\omega+kv)}{\sqrt{2}v\sqrt{\ell^2+\tau^2v^2}}
   \end{equation}
   and
   \begin{equation}
      G_{p,q}^{m,n}\left(z,r\left|\begin{matrix}a_1,\ldots,a_n,a_{n+1},\ldots,a_p\\b_1,\ldots,b_m,b_{m+1},\ldots,b_q\end{matrix}\right.\right)
         =\frac{r}{2i\pi}\int_{\gamma}\frac
         {\Gamma(1-a_1-rs)\ldots\Gamma(1-a_n-rs)\Gamma(b_1+rs)\ldots\Gamma(b_m+rs)}
         {\Gamma(a_{n+1}+rs)\ldots\Gamma(a_p+rs)\Gamma(1-b_{m+1}-rs)\ldots\Gamma(1-bq-rs)}
         z^{-s}\ud s
   \end{equation}
   is the generalized Meijer G-function, $\gamma$ being an appropriate path in the complex plane, and $\Gamma$ is the Gamma
   function. To have a practical expression of the scattering mean-free path, we use the on-shell approximation in
   Eq.~\eqref{eq:sigma_gauss}, which leads to
   \begin{multline}\label{eq:sigma_gauss_path}
      \im\Sigma(k_b,\omega)=\frac{AB\ell\tau\omega^2 v^2}{4c^4}\left\{
         \omega\sqrt{\frac{2\pi}{\mathcal{W}}}+\frac{v}{\mathcal{W}}e^{-\mathcal{Z}}
         +\frac{\omega}{\mathcal{W}}\sqrt{\frac{\pi}{2\mathcal{W}}}e^{-\mathcal{X}}
            \left(\ell^2-v^2\tau^2\right)\left[\operatorname{erf}\left(\sqrt{\mathcal{Y}}\right)-1\right]
      \right.
   \\
      \left.
         +\frac{\omega}{\mathcal{W}}\sqrt{\frac{2\pi}{\mathcal{W}}}e^{-\mathcal{X}}
            \ell^2\left[-2+Q\left(-\frac12,\mathcal{Y}\right)\right]
         -\omega\sqrt{\frac{\pi}{2\mathcal{W}}}e^{-\mathcal{X}}
            \left[-2+Q\left(\frac12,\mathcal{Y}\right)\right]
      \right\}
   \end{multline}
   where
   \begin{equation}
      \mathcal{W}=\epsilon_r\ell^2+v^2\tau^2
      \quad;\quad
      \mathcal{X}=\frac{2\ell^2\omega^2\tau^2}{\mathcal{W}}
      \quad;\quad
      \mathcal{Y}=\frac{2\ell^4\omega^2}{v^2\mathcal{W}}
      \quad;\quad
      \mathcal{Z}=\frac{2\ell^2\omega^2}{v^2}
   \end{equation}
\end{widetext}
and $Q$ is the regularized incomplete gamma function defined by $Q(a,z)=\Gamma(a,z)/\Gamma(a)$, with $\Gamma(a,z)$ the
incomplete gamma function. Equation~\eqref{eq:sigma_gauss_path} together with Eq.~\eqref{eq:scattering_mean_free_path}
provide the expression of the scattering mean-free path $\ell_s$ for a spatio-temporal gaussian-correlated disorder.
It is also interesting to extract the expression of the scattering
mean-free path in the limit where the time disorder vanishes (\ie, $\tau\to\infty$), which gives
\begin{equation}
   \frac{1}{\ell_{s,\tau\to\infty}(\omega)}=\frac{AB\ell\omega^2v^2}{2c^4}\sqrt{\frac{\pi}{2}}\left[
         1+\exp\left(-\frac{2\ell^2\omega^2}{v^2}\right)
      \right].
\end{equation}
Similarly, the expression in the limit of a vanishing space disorder (\ie, $\ell\to\infty$) is 
\begin{equation}
   \frac{1}{\ell_{s,\ell\to\infty}(\omega)}=\frac{AB\tau\omega^2 v^3}{2c^4}\sqrt{\frac{\pi}{2}}\left[
         1-\exp\left(-2\tau^2\omega^2\right)
      \right].
\end{equation}

To get an expression for the scattering mean-free time $\tau_s$ in the presence of space and time disorder, we have to
make use of Eq.~\eqref{eq:scattering_mean_free_time} together with Eq.~\eqref{eq:sigma_gauss_path} where all occurrences
of the variable $\omega$ are replaced by $kv$. The limited cases are given by
\begin{equation}
   \frac{1}{\tau_{s,\tau\to\infty}(k)}=\frac{AB\ell k^2v^5}{2c^4}\sqrt{\frac{\pi}{2}}\left[
         1+\exp\left(-2\ell^2k^2\right)
      \right]
\end{equation}
for a vanishing time disorder and
\begin{equation}
   \frac{1}{\tau_{s,\ell\to\infty}(k)}=\frac{AB\tau k^2 v^6}{2c^4}\sqrt{\frac{\pi}{2}}\left[
         1-\exp\left(-2\tau^2k^2v^2\right)
      \right]
\end{equation}
for a vanishing space disorder.

\subsection{Numerical simulations}
% --------------------------------

\begin{figure*}[!htb]
   \centering
   \includegraphics[width=\linewidth]{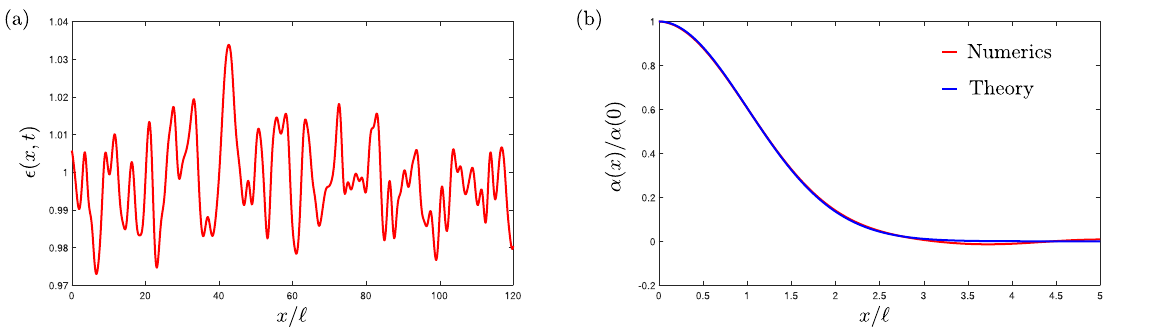}
   \caption{(a) Example of spatial disorder at a fixed time $t$. (b) Comparison between the disorder correlation
   function in space computed numerically and that given by Eq.~\eqref{eq:disorder_correlation_functions}. The parameters are:
   $\sqrt{AB}=\num{2e-2}$ and $\epsilon_b=1$. $1680$ disorder configurations are used to perform the statistical average.}
   \label{fig:disorder_correlation}
\end{figure*}

In this section we compare the predictions of the theoretical model with numerical simulations performed without
approximations.  The first step consists in generating numerically an ensemble of configurations of disorder [\ie, of
$\epsilon(x,t)$] that will be used to perfom an ensemble average. The statistics of $\epsilon(x,t)$ has to satisfy
Eq.~(\ref{eq:correlation}), which is the only assumption on the model of disorder in the theory. One way of achieving
this is to consider the particular case of a permittivity in which the space and time dependences factorize. We choose a
permittivity in the form $\epsilon(x,t)=1+\delta\epsilon(x)\delta\epsilon(t)$, with $\epsilon_b=1$ for the sake of simplicity, and $\delta\epsilon(x)$ and
$\delta\epsilon(t)$ statistically independent. In this case, we immediately find that
\begin{multline}
   \bra\delta\epsilon(x,t)\delta\epsilon(x',t')\ket=\bra\delta\epsilon(x)\delta\epsilon(t)\delta\epsilon(x')\delta\epsilon(t')\ket
\\
      =\bra\delta\epsilon(x)\delta\epsilon(x')\ket\bra\delta\epsilon(t)\delta\epsilon(t')\ket
\\
      =\alpha(x-x')\beta(t-t').
\end{multline}
Under this assumption, we only have to generate two independent one-dimensional disorders for $\delta\epsilon(x)$ and
$\delta\epsilon(t)$, with gaussian correlation functions. Let us illustrate this process with space disorder with the
correlation function $\alpha$.  We consider a finite-size medium with size $L$, and we discretize the space into $N_x$
points $x_m$ in the interval $[-L/2,L/2]$, with a step $\Delta x=x_2-x_1$. Next, we generate a white-noise gaussian
disorder [standard normal distribution $\mathcal{N}(0,1)$] that is finally convolved with
\begin{equation}
   f_m=\left(\frac{2}{\pi}\right)^{1/4}\sqrt{A\ell\Delta x}\exp\left[-\frac{x_m^2}{\ell^2}\right] \, ,
\end{equation}
which gives one realization $\delta\epsilon_m$. Restarting the process allows us to generate a set of disorder configurations.
After averaging, the correlation function tends to the function $\alpha$, as expected. The same process can be followed to generate
configurations of the time disorder, the time interval $[0,T]$ being discretized into $N_t$ points $t_n$ with a step size $\Delta t$. 
An example of disorder is plotted in Fig.~\ref{fig:disorder_correlation}\,(a)
together with a comparison between the numerical and theoretical correlation function in Fig.~\ref{fig:disorder_correlation}\,(b).

% Example of disorder and correlation function

We now briefly describe the numerical resolution of the wave equation for a given configuration of disorder. We need to solve
Eq.~\eqref{eq:wave_equation} with the boundary conditions $E(-L/2,t)=E(L/2,t)=0$, and the initial condition $E(x,0)=0$. 
The source term $S$ depends on the type of situation to be addressed. To compute the spatial evolution of the field, in order 
to estimate the scattering mean-free path, we choose
\begin{equation}
   S(x,t)=S_{\omega_0}(x,t)=s(t)\delta\left(x\right)e^{-i\omega_0 t}
\end{equation}
which corresponds to a point source oscillating at a given frequency $\omega_0$. To avoid numerical artifacts due to a
discontinuous source term in time, we apply a $C^{\infty}$ pseudo step function given by
\begin{equation}
   s(t)=\begin{cases}
      0 & \text{if $t<0$,}
   \\
      1 & \text{if $t>t_r$,}
   \\
      \exp\left\{-\left[1-(t-t_r)^2/t_r^2\right]^{-1}+1\right\}
         & \text{otherwise,}
   \end{cases}
\end{equation}
where $t_r$ is the rising time. To estimate the temporal evolution of the field, in order to compute the scattering
mean-free time, we use a source term of the form
\begin{equation}
   S(x,t)=S_{k_0}(x,t)=d(t)e^{ik_0 x}
\end{equation}
where $d$ is a $C^{\infty}$ pseudo Dirac delta function given by
\begin{equation}
   d(t)=\frac{s(t)-s(t-t_r)}{t_r},
\end{equation}
also chosen to avoid numerical artifacts. This source term corresponds to a temporal pulse oscillating in space with
a fixed wavevector $k_0$.

To solve the wave equation, we simply discretize it in space and time, with the numerical scheme
\begin{multline}
   E_{m,n+1}=\frac{2\epsilon_{m,n}E_{m,n}-\epsilon_{m,n-1}E_{m,n-1}}{\epsilon_{m,n+1}}
      +\frac{c^2\Delta t^2}{\epsilon_{m,n+1}}
\\\times
      \left[
         S_{m,n}
         +\frac{E_{m+1,n}+E_{m-1,n}-2E_{m,n}}{\Delta x^2}
      \right]
\end{multline}
where the first indices ($m-1$, $m$, $m+1$) correspond to space discretization, and the second indices ($n-1$, $n$,
$n+1$) to time discretization. The Dirac delta function in the source term is discretized using a Kronecker delta (\ie,
$\delta_{m,m_0}/\Delta x$ where $m_0$ is the index corresponding to $x=0$). As for any finite-difference scheme, the
Courant–Friedrichs–Lewy condition must be fulfilled to ensure numerical convergence and stability (\ie,
$\Delta t\le \Delta x/c$). The resolution is performed for each disorder configuration of the ensemble, allowing us to
estimate the ensemble averaged electric field.

\begin{figure}[!htb]
   \centering
   \includegraphics[width=0.8\linewidth]{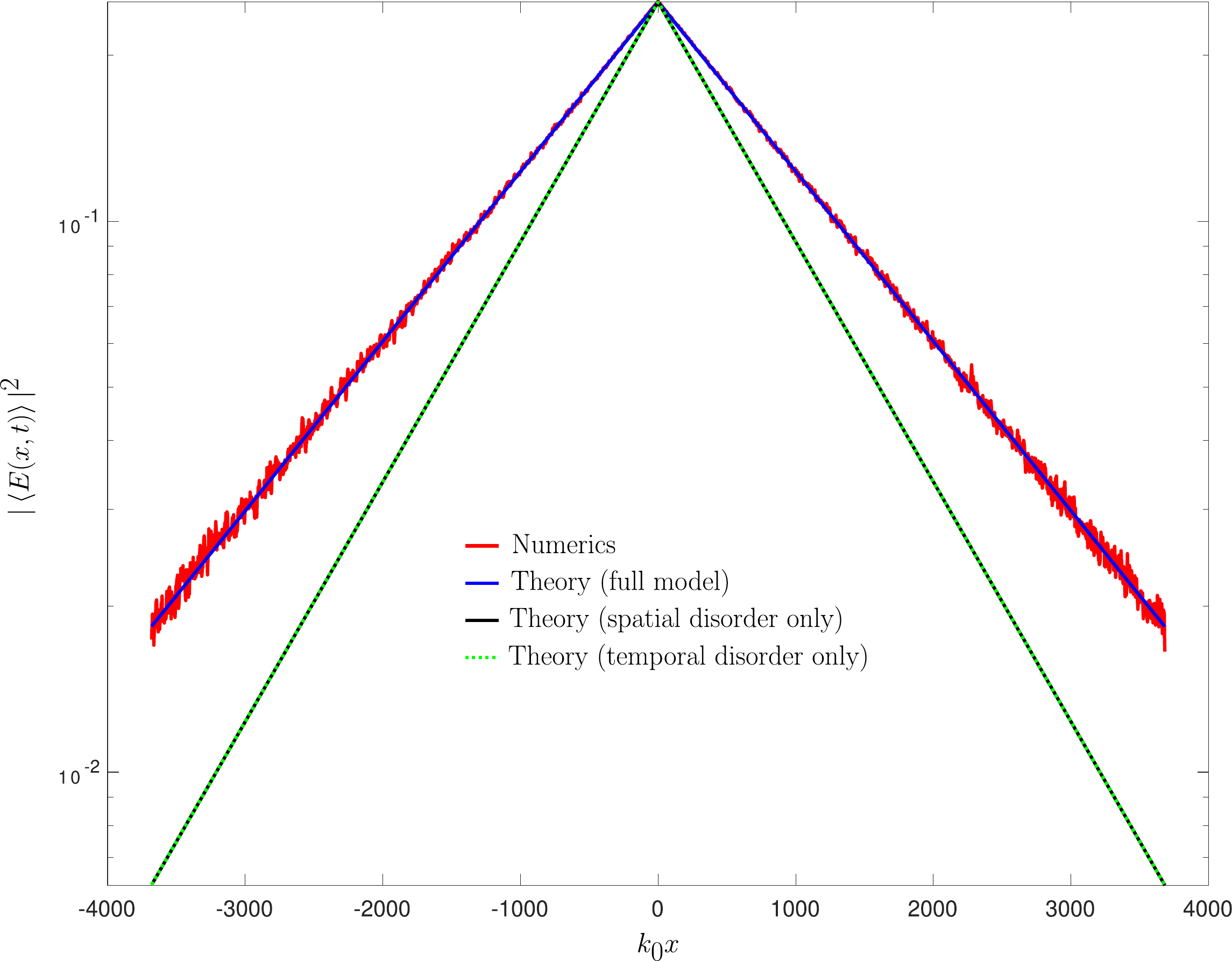}
   \caption{Intensity of the average field versus the normalized space variable $k_0x$, with $k_0=\omega_0/c$.  This
   intensity is computed numerically (red solid line) and analytically (blue solid line for the full model, black dotted
   line for the model taking into account the space disorder only, \ie $\tau\to\infty$, and green dotted line for the
   model taking into account the time disorder only, \ie $\ell\to\infty$). The plot corresponds to the normalized time
   $\omega_0t=4000$. The parameters are: $k_0L=8000$, $k_0\ell=4$,
   $\omega_0\tau=4$, $\sqrt{AB}=\num{2e-2}$, $\epsilon_b=1$ and $\omega_0t_r=100$. $10^4$ disorder
   configurations are used to perform the statistical average.}
   \label{fig:intens_space}
\end{figure}

Let us start with the spatial evolution of the field, with the source term $S_{\omega_0}$. We plot in
Fig.~\ref{fig:intens_space} the intensity of the average field obtained from the full numerical simulation and from the
analytical expressions, for the parameters given in the figure caption. The numerical result of $\left|\bra
E(x,t)\ket\right|^2$ at a fixed long time $t$ is compared to the square modulus of the average Green function (\ie,
$\left|\bra G(x,\omega_0)\ket\right|^2$ given by Eq.~\eqref{eq:intensity_spatial}, with $k_r=k_b$). Excellent
quantitative agreement is observed, which supports the validity of the theoretical model for the scattering mean-free
path $\ell_s$. We also see that taking into account the spatial disorder only does not lead to an accurate result. The
full model given by Eqs.~\eqref{eq:scattering_mean_free_path} and \eqref{eq:sigma_gauss_path} is needed to provide a
relevant prediction, showing that the time dependence of the disorder clearly affects the spatial attenuation of the field.
We also note that the scattering mean-free path is larger for the full disorder model than for spatial disorder model
only, meaning that adding time disorder reduces the effect of scattering from space disorder. This result may look
counter-intuitive, but it is because energy is not conserved in the presence of time disorder.

% Numerical simulation for the scattering mean-free path

\begin{figure}[!htb]
   \centering
   \includegraphics[width=0.8\linewidth]{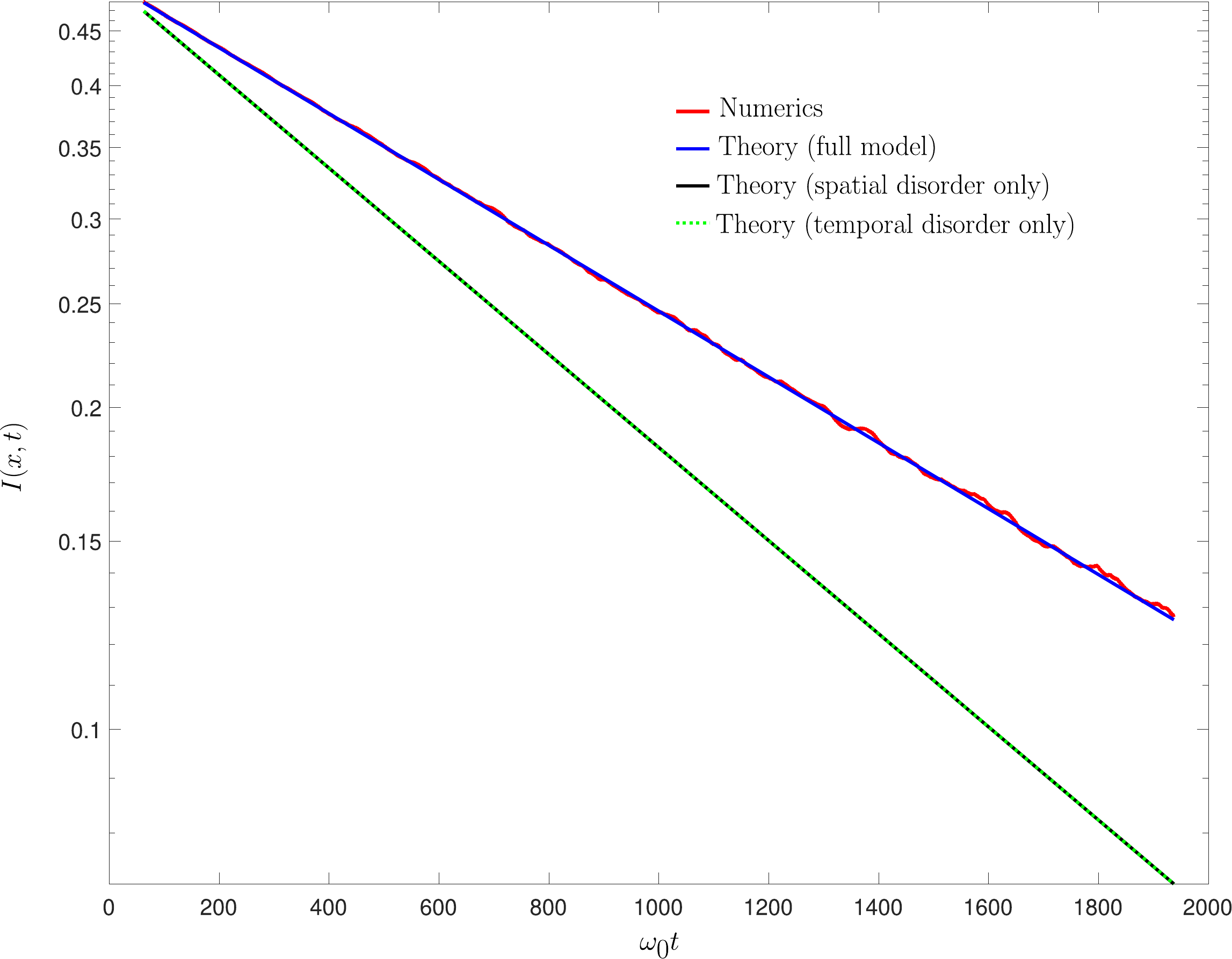}
   \caption{Intensity of the average field as a function of the normalized time variable $\omega_0t$ with
   $\omega_0=k_0c$.  This intensity is computed numerically (red solid line) by applying
   Eq.~\eqref{eq:intens_no_oscillations} and analytically (blue solid line for the full model, black dotted line for the
   model taking into account the space disorder only, \ie $\tau\to\infty$, and green dotted line for the model taking
   into account the time disorder only, \ie $\ell\to\infty$). The plot corresponds to a fixed normalized $k_0x=0$. The
   parameters are: $k_0L=8000$, $k_0\ell=4$, $\omega_0\tau=4$,
   $\sqrt{AB}=\num{2e-2}$, $\epsilon_b=1$, $\omega_0t_r=0.1$ and $\omega_0T=30$. $10^4$ disorder
   configurations are used to perform the statistical average. Short times are not represented since for $t<T$,
   the averaging procedure given by Eq.~\eqref{eq:intens_no_oscillations} leads to an oscillating signal because of the
   Heaviside step function at $t=0$.}
   \label{fig:intens_time}
\end{figure}

Next, we study the temporal evolution of the field with the source term $S_{k_0}$. In order to compare the numerical
results to the square of the average Green function (\ie, $\overline{\bra G(k_0,t)\ket^2}$, with
$\omega_r=\omega_b$), we first compute numerically the average field $\bra E(x,t)\ket$ for a fixed $x=0$. Then, we take
the square modulus and perform a rolling average over a time window with width $T$ satisfying $2\pi/\omega_b\ll
T\ll \tau_s$. This eliminates rapid oscillations and keeps the decaying envelope that depends on the scattering
mean-free time $\tau_s$. We obtain
\begin{equation}\label{eq:intens_no_oscillations}
   I(x,t)=\int_{-\infty}^{+\infty} w(t-t')\left|\bra E(x,t')\ket\right|^2\ud t'
\end{equation}
where $w$ is a rectangular function of width $T$ and amplitude $1/T$. The comparison is plotted in
Fig.~\ref{fig:intens_time}. Again, we obtain excellent quantitative agreement between the numerical simulation and the
analytic expressions, supporting the theoretical model for the scattering mean-free time $\tau_s$. We also observe that
the full theoretical model taking into account both space and time disorder is required to correctly predict the time
decay of the intensity.

% Numerical simulation for the scattering mean-free time

\section{Conclusion}
% ==================

In conclusion, we have studied the behavior in space and time of the averaged field propagating in a medium with both
space and time disorders. We have developed a multiple scattering theory that predicts the space and time decay of the
average field, and allows to derive practical expressions of the scattering mean-free path $\ell_s$ and mean-free time
$\tau_s$ in the weak-scattering regime. The model has been compared to exact numerical simulations, showing quantitative
agreement in the particular case of a spatio-temporal gaussian-correlated disorder with no space-time cross
correlation.  Counter-intuitively, in this regime the introduction of a time disorder on top of a space disorder tends
to reduce the scattering strength, even in the absence of cross correlation between the two types of disorders.
However, this theory does not seem to predict the existence of a situation where the attenuation by scattering is
totally cancelled out, at least in the case of Gaussian-correlated disorder. The theory developed in this work and the
results bring a brick in the widely open field of waves in complex space and time varying media. In particular, the
next step will be to study the case of the average intensity, whose behavior is known to be very different from that of
the average electric field in the presence of disorder.

\section*{Aknowledgments}
% =======================

This work has received support under the program ``Investissements d’Avenir'' launched by the French Government.

\section*{Disclosures}
% ====================

The authors declare no conflicts of interest.

% =======
\appendix
% =======

\section{Calculation of the Green function $G_b$}\label{green_calculation}
% =================================================

% Contours
\begin{figure*}[!htb]
   \centering
   \includegraphics[width=\linewidth]{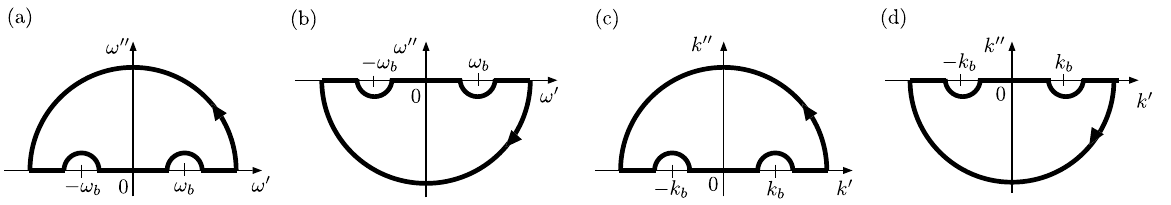}
   \caption{Various integration contours used to compute the Green function of the reference medium. (a) and (b) are
   used for the inverse Fourier transform for negative and positive times respectively. (c) and (d) are used for the
   inverse Fourier transform for positive and negative positions respectively.}
   \label{fig:contours}
\end{figure*}

The 1D scalar Green function of the wave equation in the reference medium described by its relative permittivity
$\epsilon_b$ is given by Eq.~\eqref{eq:wave_equation_reference_green} which reads in the Fourier domain
\begin{equation}\label{eq:wave_equation_reference_green_fourier}
   \left(k^2-\frac{\omega^2}{v^2}\right)G_b(k,\omega)=1
\end{equation}
where we recall that $v=c/\sqrt{\epsilon_b}$. $k$ and $\omega$ are the dual variables for $x$ and $t$ respectively. In
terms of distributions (the Green function is rigorously a distribution), the inversion of this equation leads to
\begin{equation}\label{eq:green_fourier_unknowns}
   G_b(k,\omega)=\operatorname{PV}\left[\frac{1}{k^2-\omega^2/v^2}\right]
      +\lambda\delta\left(k-\frac{\omega}{v}\right)
      +\mu\delta\left(k+\frac{\omega}{v}\right)
\end{equation}
where $\lambda$ and $\mu$ are constants that should be determined in order to fulfil the boundary conditions in space
and time. For that purpose, we consider first the inverse Fourier transform in time which gives
\begin{multline}
   G_b(k,t)=\operatorname{PV}\int_{-\infty}^{+\infty}\frac{e^{-i\omega t}}{\omega_b^2/v^2-\omega^2/v^2}\frac{\ud\omega}{2\pi}
      +\frac{\lambda v}{2\pi}e^{-ikvt}
\\
      +\frac{\mu v}{2\pi}e^{ikvt}
\end{multline}
where we recall that $\omega_b=kv$. To compute the first term, we apply the residue theorem and consider two cases. If
$t<0$, we use the contour described in Fig.~\ref{fig:contours}\,(a). The semicircle in the upper plane is chosen in
order to apply Jordan's lemma. This leads to
\begin{equation}
   G_b(k,t<0)=-\frac{v^2}{2\omega_b}\sin(\omega_b t)
      +\frac{\lambda v}{2\pi}e^{-i\omega_b t}
      +\frac{\mu v}{2\pi}e^{i\omega_b t}.
\end{equation}
The causality requires that $G_b(k,t<0)=0$ which leads to $-\lambda=\mu=\pi/(2ik)$. For $t>0$, we use the contour
described in Fig.~\ref{fig:contours}\,(b). This finally gives
\begin{equation}\label{eq:green_time}
   G_b(k,t)=\frac{\operatorname{H}(t)v^2}{\omega_b}\sin\left(\omega_b t\right).
\end{equation}
We consider now the inverse Fourier transform in space with the values of $\lambda$ and $\mu$ determined above. This
gives
\begin{equation}
   G_b(x,\omega)=\operatorname{PV}\int_{-\infty}^{+\infty}\frac{e^{ikx}}{k^2-k_b^2}\frac{\ud k}{2\pi}
      +\frac{i}{2k_b}\cos\left(k_bx\right)
\end{equation}
where we recall that $k_b=\omega/v$. Again, we consider two cases to compute the first term. If $x>0$, we use the
contour described in Fig.~\ref{fig:contours}\,(c). This leads to
\begin{multline}
   G_b(x>0,\omega)=\frac{i}{2k_b}\left[\cos\left(k_bx\right)+i\sin\left(k_bx\right)\right]
\\
      =\frac{i}{2k_b}\exp\left(ik_bx\right).
\end{multline}
If $x<0$, we use the contour described in Fig.~\ref{fig:contours}\,(d) which gives
\begin{multline}
   G_b(x<0,\omega)=\frac{i}{2k_b}\left[\cos\left(k_bx\right)-i\sin\left(k_bx\right)\right]
\\
      =\frac{i}{2k_b}\exp\left(-ik_bx\right).
\end{multline}
The two previous results combined give
\begin{equation}\label{eq:green_space}
   G_b(x,\omega)=\frac{i}{2k_b}\exp\left(ik_b|x|\right).
\end{equation}
It is interested to note that this last expression automatically fulfiled the outgoing wave condition thanks to
causality.

% Biblio

\end{document}